\documentclass[numreferences]{kluwer}

\usepackage{graphicx}
\usepackage{dcolumn}
\usepackage{bm}
\usepackage{bbm}
\usepackage{mathrsfs}
\usepackage{psfrag}
\usepackage{epsfig}
\usepackage{color}
\usepackage{fancyheadings}
\usepackage{mathbbol}
\newcommand{\nn}{\nonumber}
\newcommand{\bra}{\langle}
\newcommand{\ve}{\vert}
\newcommand{\ket}{\rangle}

\begin{document}
\begin{article}
\begin{opening}


\title{Gaussian quantum fluctuations in interacting many particle systems}
\subtitle{A Lyapunov type central limit theorem for mixing quantum systems}

\author{Michael \surname{Hartmann}\email{michael.hartmann@dlr.de}} 
\institute{Institute of Thechnical Physics, DLR Stuttgart and\\
Institute of Theoretical Physics I, University of Stuttgart}
\author{G\"unter \surname{Mahler}}
\institute{Institute of Theoretical Physics I, University of Stuttgart}
\author{Ortwin \surname{Hess}}
\institute{Advanced Technology Institute, University of Surrey}

\date{\today}

\begin{abstract}
We consider a many particle quantum system, in which each particle interacts only with its nearest
neighbours. Provided that the energy per particle has an upper bound, we show, that the energy
distribution of almost every product state becomes a Gaussian normal distribution
in the limit of infinite number of particles. We indicate some possible applications.
\end{abstract}

   

\end{opening}
%
%

Physical systems, composed of interacting identical (or similar) subsystems
appear in many branches of physics. They are standard
in condensed matter physics.

Assuming that each subsystem only interacts with its nearest neighbours
and that the energy per subsystem has an upper limit, which must not depend on the number
of subsystems $n$, we show that the distribution of energy eigenvalues of almost every product state
converges to the Gaussian normal distribution in the limit of infinitely many subsystems.
To the best of our knowledge, this fundamental quantum feature has not yet been recognized in the
literature \cite{Lieb1966,Thirring2002,Lieb2001}.

Central limit theorems for the distribution of energy eigenvalues in
quantum gases with Boltzmann statistics \cite{Goldstein1985} as well as for Bose and Fermi
statistics \cite{Goldstein1985a} have been discussed by M. Sh. Goldstein.
His theorems apply for mixed states, namely classical mixtures of quantum states involving
classical probabilities.

We consider here the distribution of energy eigenvalues for a pure quantum state.
In our case, the distribution is thus of purely quantum nature,
it solely exists because the state we consider is not an eigenstate of the
energy operator of our system.
Our theorem may be viewed as a central limit theorem for mixing quantum systems.

Some extensions of the central limit theorem to quantum systems, with the state not necessarily
being mixed, have been proven in the past 
\cite{Cushen1971,Accardi1985,Verbeure1989,Goderis1989,Kuperberg2002,Michoel2003}.
The version, which appears closest related to ours, has been published by Goderis and Vets in 1989
\cite{Goderis1989}. They consider a quantum lattice system and assume that the state and the operator,
they look at, are invariant under lattice translations. Their proof is then based on a set of
``cluster conditions'', which replace the mixing condition of the random variable case.

The assumptions we use are stricter with respect to the mixing behavior, nevertheless, they may still
be weakened and generalised.
On the other hand, we do not assume translational invariance of the operators or the state.
Instead, we use a quantum analogue of the Lyapunov condition for random variables.
This generalisation opens up a large field of applications.
The version of Coderis and Vets only applies to products of identical subsystem states, while ours
applies to almost every product state with the fraction of exceptions being negligible.

Knowing the energy distribution of a product state one can
deduce estimates on various quantities of interest.
If the system is known to be in a product state at some initial time,
one can calculate it's energy distribution.
As this distribution is conserved under Schr\"odinger dynamics, one can then make
predictions on the dynamics of the state, even in the long time limit.
These circumstances should prove helpful in many problems related to chaotic or non-chaotic
behavior in quantum systems.

On the other hand, for a given global state of the total system being a function of the total
Hamiltonian $H$, one
can calculate occupation probabilities of product states. Since only in the product basis traces
over single subsystems can be performed, this procedure allows to calculate properties of the reduced
density matrix of any subsystem. In this way, one could analyse the local properies of stationary
global states.

Finally, our theorem may underly the often-used assumption of Gaussian fluctuations.

\textbf{Notation:}
We first give some notation and definitions we use:
We consider a chain of quantum systems with next neighbour interactions.
Let the entire system be described by a Hamiltonian $H$ which is a linear, self-adjoint operator on
a seperable, complex Hilbert space $\mathscr{H}$.
The Hilbert space $\mathscr{H}$ is a direct product of the Hilbert spaces of the subsystems,
\begin{equation}\label{hilbert}
\mathscr{H} \equiv  \prod_{\mu = 1}^n \otimes \mathscr{H}_{\mu},
\end{equation}
and the Hamiltonian may be written in the form,
\begin{equation}\label{hamil}
H \equiv \sum_{\mu = 1}^n \mathcal{H}_{\mu},
\end{equation}
with
\begin{equation}
\mathcal{H}_{\mu} \equiv \I^{\otimes \mu-1} \otimes H_{\mu} \otimes \I^{\otimes n - \mu} \: + \:
\I^{\otimes \mu-1} \otimes I_{\mu,\mu+1} \otimes \I^{\otimes n - (\mu + 1)},
\end{equation}
where $H_{\mu}$ is the proper  Hamiltonian of subsystem $\mu$,
and $I_{\mu,\mu+1}$ the interaction of subsystem $\mu$ with subsystem $\mu+1$.
$\I$ is the identity operator. We chose the boundary condition $I_{n,n+1} = 0$.

Let $E_{\varphi}$ be the eigenenergies and, using the Dirac notation \cite{Sakurai}, let
$\{ \ve \varphi \ket \}$ be an orthonormal basis of $\mathscr{H}$
consisting of eigenstates of the total system.
\begin{equation}
H \ve \varphi \ket = E_{\varphi} \ve \varphi \ket \enspace \textrm{with} \enspace \enspace
\bra \varphi \ve \varphi' \ket = \delta_{\varphi \varphi'} \, ,
\end{equation}
where $\delta_{\varphi \varphi'}$ is the Kronecker delta.
We denote by $\ve a \ket$ the product state
\begin{equation}\label{prod}
\ve a \ket \equiv \prod_{\mu = 1}^n \otimes \, \ve a_{\mu} \ket,
\end{equation}
built up from some state $\ve a_{\mu} \ket$ of each subsystem $\mu$, $\ve a_{\mu} \ket \in \mathscr{H}_{\mu}$.

We furthermore define,
\begin{eqnarray}
\overline{E}_a & \equiv & \bra a \ve H \ve a \ket \\
\sigma_a^2 & \equiv & \bra a \ve H^2 \ve a \ket - \bra a \ve H \ve a \ket^2,
\end{eqnarray}
and introduce the operator
\begin{equation}
Z_n \equiv \frac{H - \overline{E}_a}{\sigma_a}
\end{equation}
which is diagonal in the same basis as $H$. Let $z_{\varphi}$ denote its eigenvalues,
\begin{equation}
Z_n \ve \varphi \ket = z_{\varphi} \ve \varphi \ket.
\end{equation}
Note that $H$ and therefore $\overline{E}_a$, $\sigma_a$ and $z_{\varphi}$ as well as the basis
$\{ \ve \varphi \ket \}$ depend on $n$.

Since $H$ and thus $Z_n$ are self-adjoint, $\ve a \ket$ induces a measure on the spectrum of $Z_n$
respective $H$. This measure of the
quantum mechanical distribution of the eigenvalues of $Z_n$ in the state $\ve a \ket$, is given by
the usual formula,
\begin{equation}
\mathbbm{P}_a \left( z_{\varphi} \in \left[z_1, z_2 \right] \right) = 
\sum_{\{ \ve \varphi \ket : z_1 \le z_{\varphi} \le z_2 \}} \ve \bra a \ve \varphi \ket \ve^2,
\end{equation}
where the sum extends over all states $\ve \varphi \ket$ with eigenvalues in the respective interval.
%
%

\textbf{Theorem:}
If the operator $H$ and the state $\ve a \ket$ satisfy
\begin{equation} \label{vacuumfluc}
\sigma_a^2 \ge n \, C
\end{equation}
for all $n$ and some $C > 0$ and if each operator $\mathcal{H}_{\mu}$ is bounded, i.e.
\begin{equation} \label{bounded}
\bra \chi \ve \mathcal{H}_{\mu} \ve \chi \ket \le C'
\end{equation}
for all normalised states $\ve \chi \ket \in \mathscr{H}$ and some constant $C'$,
then the quantum mechanical distribution of the eigenvalues of $Z_n$ in the state $\ve a \ket$ converges
weakly to a Gaussian normal distribution:
\begin{equation}\label{limit}
\lim_{n \rightarrow \infty} \mathbbm{P}_a \left( z_{\varphi} \in \left[z_1, z_2 \right] \right) =
\int_{z_1}^{z_2} \frac{\exp \left( - \, z^2 / 2 \right)}{\sqrt{2 \pi}} \, dz
\end{equation}
for all $- \infty < z_1 <z_2 < \infty$.
%
%

\textbf{Proof:}
Following the proof of the central limit theorem for mixing sequences \cite{Linnik1971} as a guideline,
we prove the statement (\ref{limit}) in three steps: First, we show that the characteristic function of $H$
does not change if a few of the $\mathcal{H}_{\mu}$ are neglected. Second, we prove, that the
characteristic function of the remainder of $H$ factorises. In the last step, we then show that
the condition for Lyapunov's version of the central limit theorem is fulfilled for
the remainder of $H$. The proof is then completed by the standard proof of 
Lyapunov's central limit theorem, which can be found in several textbooks \cite{Billingsley1995}.

Define the operators
$X_{\mu} \equiv \mathcal{H}_{\mu} - \bra a \ve \mathcal{H}_{\mu} \ve a \ket$
and split the sum
\begin{equation}
Z_n = \frac{1}{\sigma_a} \sum_{\mu = 1}^n X_{\mu}
\end{equation}
into alternate blocks of length $k-1$ (large blocks) and of length $1$ (small blocks).
The large blocks are given by
\begin{eqnarray}\label{bigblocks}
\xi_j & = & \sum_{l = 1}^{k-1} X_{(j - 1) \cdot k + l} \enspace \enspace \textrm{for}
\enspace \enspace j = 1, \dots, \left[n / k \right] \enspace \enspace \textrm{and}\\
\xi_{[n/k] + 1} & = & \sum_{l = 1}^{q} X_{[n/k] \cdot k + l} \enspace \enspace \textrm{with}
\enspace \enspace q = n - k \, [n/k],
\end{eqnarray}
where $[x]$ means the integer part of $x$
and the small blocks are the $X_{j \cdot k}$ with $j = 1, \dots, \left[n / k \right]$.
Sum up all large blocks and all small blocks separately,
\begin{equation}
\label{split}
Z_n' = \frac{1}{\sigma_a} \, \sum_{j = 1}^{[n/k] + 1} \xi_j \enspace \enspace \textrm{and}
\enspace \enspace Z_n'' = \frac{1}{\sigma_a} \, \sum_{j = 1}^{[n/k]} X_{j \cdot k},
\end{equation}
so that $Z_n = Z_n' + Z_n''$.
The integer block length $k$ is chosen to depend on $n$ ($k = k(n)$) such that
\begin{equation}
\lim_{n \to \infty} \, \frac{n}{k^2} = 0 \enspace \enspace \textrm{and} \enspace \enspace
\lim_{n \to \infty} \, \frac{k}{n} = 0,
\end{equation}
with $k = \left[ n^{3/4} \right]$ being a possible realisation.

Consider the characteristic function
\begin{equation}
\bra a \ve e^{-i r Z_n} \ve a \ket
\end{equation}
with real $r$.

First let us show that for all real $r$ and $n \rightarrow \infty$:
\begin{equation}
\label{cfconv}
\bra a \ve e^{-i r Z_n} \ve a \ket \rightarrow \bra a \ve e^{-i r Z_n'} \ve a \ket
\end{equation}
Using the operator identity \cite{Fick1990}
\begin{equation}
e^{-i r (A + B)} = e^{-i r A} - i \int_{0}^{r} e^{-i (r - s) (A + B)} B e^{-i s A} ds,
\end{equation}
the triangle- and the Schwarz-inequality, one gets
\begin{eqnarray}
\ve \bra a \ve e^{-i r Z_n} - e^{-i r Z_n'} \ve a \ket \ve & \le &
\int_{0}^{r} ds \sqrt{\bra a \ve e^{i s Z_n} \left(Z_n''\right)^2 e^{-i s Z_n} \ve a \ket} \nn \\
& \le & r \, \sqrt{\left(\frac{1}{n} \left[ \frac{n}{k} \right]^2 \right)\frac{(2 C')^2}{C}}
\end{eqnarray}
which, indeed, converges to zero for $n \rightarrow \infty$.

Next, we show that the characteristic function of $Z_n'$ factorises.
\begin{equation}
\label{factor}
\bra a \ve e^{-i r Z_n'} \ve a \ket = \prod_{j=1}^{[n/k] + 1} \bra a \ve e^{-i r \xi_j} \ve a \ket
\end{equation}
To this end, we first note two
important properties that arise due to the next neighbour interaction and the product property
of the state $\ve a \ket$: For $\ve \mu - \nu \ve > 1$ and any two integers $k$ and $l$, we have
\begin{eqnarray}
\left[ \mathcal{H}_{\mu} , \mathcal{H}_{\nu} \right] & = & 0 \label{commut_ha_mu}\\
\bra a \ve \left( \mathcal{H}_{\mu} \right)^k
\left( \mathcal{H}_{\nu} \right)^l \ve a \ket & = &
\bra a \ve \left( \mathcal{H}_{\mu} \right)^k \ve a \ket \, 
\bra a \ve \left( \mathcal{H}_{\nu} \right)^l \ve a \ket \label{factor_ha_mu}.
\end{eqnarray}
Therefore, for all $(i,j)$ and any two integers $k$ and $l$,
\begin{eqnarray}
\left[ \xi_i, \xi_j \right] & = & 0 \label{commutexi}\\
\bra a \ve \left(\xi_i \right)^k \left(\xi_j \right)^l \ve a \ket & = &
\bra a \ve \left( \xi_i \right)^k \ve a \ket \, 
\bra a \ve \left( \xi_j \right)^l \ve a \ket, \label{factorisexi}
\end{eqnarray}
and equation (\ref{factor}) follows as a direct consequence.

Finally we prove that the $\xi_j$ fulfill the Lyapunov condition:
\begin{equation}
\label{lyapunov}
\lim_{n \to \infty} \frac{1}{\sigma_a^{2+m}} \sum_{j=1}^{[n/k] + 1}
\bra a \ve \, \ve \xi_j \ve^{2+m} \, \ve a \ket = 0
\end{equation}
for some $m > 0$.
Note that due to equation (\ref{cfconv}),
$\bra a \ve \left( Z_n' \right)^2 \ve a \ket \rightarrow \sigma_a^2$ as $n \rightarrow \infty$ and
therefore equation (\ref{lyapunov}) is, indeed, the Lyapunov condition for the $\xi_j$.
We verify the condition for $m = 2$. To this end, consider
\begin{equation}
\bra a \ve \xi_j^4 \ve a \ket = \sum_{\mu, \nu, \rho, \tau = 1}^{k-1}
\bra a \ve X_{(j-1) k + \mu} \, X_{(j-1) k + \nu} \, X_{(j-1) k + \rho} \, X_{(j-1) k + \tau} \ve a \ket.
\end{equation}
Since $\bra a \ve X_{\mu} \ve a \ket = 0$ and because of
equations (\ref{commutexi}) and (\ref{factorisexi}), only those terms are nonzero,
for which all the $X_{\mu}$
are identical or neighbours or where two pairs of identical or neighbouring $X_{\mu}$ appear.
For example $\bra a \ve X_{\mu} X_{\mu+1} X_{\mu+2} X_{\mu-1} \ve a \ket \not= 0$ while
$\bra a \ve X_{\mu} X_{\mu+1} X_{\mu+3} X_{\mu-1} \ve a \ket = 0$ or
$\bra a \ve X_{\mu} X_{\mu+1} X_{\nu-1} X_{\nu} \ve a \ket \not= 0$ while
$\bra a \ve X_{\mu} X_{\nu} X_{\mu} X_{\nu+2} \ve a \ket = 0$.
Using this fact and the conditions (\ref{vacuumfluc}) and (\ref{bounded}) one realises that
\begin{eqnarray}
& & \frac{1}{\sigma_a^4} \sum_{j=1}^{[n/k] + 1} \bra a \ve \xi_j^4 \ve a \ket \le \nn \\
& & \le \left( \left[\frac{n}{k}\right] + 1 \right)
\frac{\left( (k-1)^2 + 3 \cdot 5 \cdot 7 \cdot 3! \cdot (k-1)\right) (2 C')^4}{n^2 C^2}
\end{eqnarray}
where the rhs vanishes in the limit $n \rightarrow \infty$. Note that
$\bra a \ve \xi_{[n/k]+1}^4 \ve a \ket$ contains less terms than $\bra a \ve \xi_j^4 \ve a \ket$ for
$j < [n/k]+1$ and is therefore bounded by the same expression. 

With the arguments above showing that the characteristic function of $Z_n'$ factorises and that the
$\xi_j$ obey the Lyapunov condition, it is straight forward to prove, following the standard steps
\cite{Billingsley1995}, that
\begin{equation}
\lim_{n \to \infty} \, \bra a \ve \exp \left(- i \, r \, Z_n' \right) \ve a \ket =
\exp \left(- \, \frac{r^2}{2}\right)
\end{equation}
and, using equation (\ref{cfconv}) one concludes that
\begin{equation} \label{gauss_ch_f}
\lim_{n \to \infty} \,
\bra a \ve \exp \left( - i \, r \, Z_n \right) \ve a \ket = \exp \left(- \, \frac{r^2}{2} \right).
\end{equation}
Here both limits are pointwise for all real $r$.
 
Finally, the continuity theorem \cite{Billingsley1995} states that the
pointwise convergence of the characteristic functions, established above,
implies the weak convergence of the distributions.
The density of the limit distribution is thus given by the Fourier transform of the
characteristic function in equation (\ref{gauss_ch_f}), which proves our theorem.

Note that all the steps not explicitely carried out here only use properties of Lebesgue integration
and no further properties of probability distributions. We therefore do not run into
difficulties related to so called ''no hidden variable theorems'' \cite{Duerr1992,Duerr2003}.

For applications in physics, where $n$ is very large but finite, the density of the
limit distribution can be written as a function of the energy $E$ of the system,
\begin{equation} \label{limit_phys}
\rho_a (E) =
\frac{1}{\sqrt{2 \pi} \, \sigma_a} \: \exp \left( - \frac{\left(E - \overline{E}_a \right)^2}
{2 \, \sigma_a^2} \right),
\end{equation}
so that $\mathbbm{P}_a (E \in [E_1,E_2]) = \int_{E_1}^{E_2} \rho_a (E) dE$.
%
%

\textbf{Discussion and Generalisations:}
Let us first analyse the conditions (\ref{vacuumfluc}) and (\ref{bounded}) in more detail.

Rewriting (\ref{vacuumfluc}) in terms of the
operators $X_{\mu}$ and using equation (\ref{factor_ha_mu}) we get
\begin{equation}
\label{vacuumfluc2}
\sum_{\mu=1}^n \bra a \ve \frac{1}{2} \left( X_{\mu}^2 + X_{\mu+1}^2 \right) +
X_{\mu} X_{\mu+1} + X_{\mu+1} X_{\mu} \ve a \ket \ge n C.
\end{equation}
Thus, every term in the sum in (\ref{vacuumfluc2}) being larger than $C$ is sufficient for
(\ref{vacuumfluc}) to be satisfied.

Condition (\ref{bounded}) physically states, that the excitation energy must not be concentrated in only a
small part of the subsystems.
For very large systems, where our theorem applies, this is only a minor restriction since the fraction
of states that do not fulfill condition (\ref{bounded}) is vanishingly small.

Several conditions we have used to derive our theorem may be relaxed and substituted by weaker assumptions.

First of all, the theorem is not only valid for a linear chain but also for
two and three dimensional lattices. 

It is also straight forward to proof the same theorem for periodic boundary conditions, $I_{n,n+1} = I_{n,1}$.

In addition, it is obviously not necessary that the subsystems only interact with their nearest neighbours.
The theorem holds as long as the number of interaction partners (the connectivity)
of each particle is limited.

Furthermore, the observable one considers need not be the Hamiltonian. Any other observable
shows the same feature as long as conditions
(\ref{vacuumfluc}), (\ref{bounded}), (\ref{commut_ha_mu}) and (\ref{factor_ha_mu}) are met.

Let us stress here, that neither the operator (Hamiltonian) nor the product state need to be invariant
with respect to lattice translations.

Finally, conditions (\ref{vacuumfluc}) and (\ref{bounded}) may be relaxed, since the theorem
still holds, whenever Lyapunov's condition, or even only Lindeberg's condition \cite{Billingsley1995},
is fulfilled. We have chosen here stricter but simpler conditions to make it easier to check
the applicability of our theorem.
%
%

\textbf{Examples:}
We mention two examples, where our theorem applies.
First, we consider an Ising spin chain of the type \cite{Mahler1998}
\begin{equation}
H = - B \sum_i s_i^z - \frac{J}{2} \sum_i s_i^x \otimes s_{i+1}^x.
\end{equation}
Here, $s_i^x$ and $s_i^z$ are the Pauli matrices, 
\begin{equation}
s^x = \left( \begin{array}{cc} 0 & 1 \\ 1 & 0 \end{array} \right) \enspace \enspace \textrm{and}
\enspace \enspace s^z = \left( \begin{array}{cc} 1 & 0 \\ 0 & -1 \end{array} \right),
\end{equation}
$B$ the difference between local energy levels
and $J$ the coupling strength. The energy of each spin is at least $- B$ and at most $B$
so that condition (\ref{bounded}) is satisfied. The squared width $\sigma_a^2$ reads
\begin{equation}
\sigma_a^2 = n \, \frac{J^2}{4},
\end{equation}
where $n$ is the number of spins, and condition (\ref{vacuumfluc}) is also met.

As a second example, we consider a harmonic chain.
\begin{equation}
H = - \sum_i \frac{p_i^2}{2 m} + \frac{m}{2} \omega^2 \left(q_{i+1} - q_i \right)^2
\end{equation}
where $q_i$ and $p_i$ are the position and momentum of particle number $i$.
All particles have mass $m$ and the coupling has frequency $\omega$.
Since the energy of a harmonic oscillator is not bounded, our theorem only applies
to states where the energy per oscillator does not exceed a certain bound, which
on the other hand may be chosen arbitrarily large.
The squared width $\sigma_a^2$ for the harmonic chain is
\begin{equation}
\sigma_a^2 = n \, \omega^2 \, \left(n_i + \frac{1}{2}\right) \left(n_{i+1} + \frac{1}{2}\right)
\end{equation}
where $n_i$ is the occupation number of oscillator number $i$. Since $n_i \ge 0$,
condition (\ref{vacuumfluc}) is satisfied.
%
%

\textbf{Applications:}
We finally discuss two areas of possible applications of our result.

\textbf{a:} One may consider the product state $\ve a \ket$ as an initial state and make
predictions about its dynamics. Using equation (\ref{gauss_ch_f}), one can calculate the fidelity of
state $\ve a \ket$:
\begin{equation}
\left| \bra a \ve e^{-i H t} \ve a \ket \right|^2 = e^{- \sigma_a^2 t^2}
\end{equation}
Furthermore, one can give an upper bound to the transition probability to another
product state $\ve b \ket$ ($\bra a \ve b \ket = 0$) for all times $t$;
\begin{equation}
\left| \bra b \ve e^{-i H t} \ve a \ket \right|^2 \le
\frac{2 \, \sigma_a \, \sigma_b}{ \sigma_a^2 + \sigma_b^2} \:
\exp \left( - \frac{\left( \overline{E}_a - \overline{E}_b \right)^2}{2 (\sigma_a^2 + \sigma_b^2)} \right)
\end{equation}
where we have assumed $\sigma_a \ll \overline{E}_a$ and $\sigma_b \ll \overline{E}_b$ and the
ground state energy has been chosen to be zero ($E_0 = 0$).

\textbf{b:} One can calculate diagonal elements of the reduced density matrix of a selected
subsystem in the basis $\ve a_{\mu} \ket$, provided the total system is in a stationary state,
that is, its density matrix $\rho_{total}$ is a function of $H$ \cite{Jordan2003}. For this applications,
it is most interesting to take $\ve a \ket$ to be an eigenstate of the Hamilton operator
without the nearest neighbour interactions ($H_0 \ve a \ket = E_a \ve a \ket$, $H_0 = \sum_{\mu} H_{\mu}$).
Here, very interesting conclusions can be drawn on the minimal spatial extension of
temperature, which will be presented elsewhere \cite{Hartmann2003}.

In summary, we have considered a large quantum system composed of subsystems,
where each subsytem only interacts with a limited number of neighbours.
We have shown that for almost every product state,
the distribution of the total energy converges to a Gaussian normal distribution in the limit
of infinitely many subsystems (\ref{limit}).
This is the main result of this paper. The assumptions we have made are quantum mechanical analogues to the
conditions for Lyapunov's central limit theorem for mixing random variables.
We did not dwell on the most general assumptions needed for our theorem
to make the verification of our conditions in physical applications straightforward.
Nevertheless we have discussed possible generalisations as well as some preliminary applications.

We thank J.\ Gemmer, M.\ Michel, H.\ Schmidt, M.\ Stollsteimer, F.\ Tonner, M.\ Henrich and C.\ Kostoglou
for fruitful discussions.

M.H. in particular wants to thank Prof. Detlef D\"urr for many helpfull comments.

\end{article}

\begin{thebibliography}{99}


\bibitem{Lieb1966}
Lieb, E. and Mattis, D.: 1966,
'Mathematical Physics in One Dimension',
Academic Press, New York

\bibitem{Thirring2002}
Thirring W.: 2002,
'Quantum Mathematical Physics'
Springer , Berlin, 2nd ed.

\bibitem{Lieb2001}
Lieb, E.: 2001,
'The Stability of Matter: From Atoms to Stars',
Springer , Berlin

\bibitem{Goldstein1985}
Goldstein M. Sh.: 1985
{\it J. Stat. Phys.} {\bf 40},
pp.~329

\bibitem{Goldstein1985a}
Goldstein M. Sh.: 1985
{\it Theoret. and Math. Phys.} {\bf 1},
pp.~412

\bibitem{Cushen1971}
Cushen, C.D. and Hudson, R.L.: 1971
{\it J. Appl. Probability} {\bf 8},
pp.~454

\bibitem{Accardi1985}
Accardi, L. and Bach, A.: 1985,
{\it Z. Wahrsch. verw. Geb.} {\bf 68},
pp.~393

\bibitem{Verbeure1989}
Goderis, D., Verbeure, A. and Vets, P.: 1989,
{\it Prob. Th. Rel. Fields} {\bf 82},
pp.~527

\bibitem{Goderis1989}
Goderis, D. and Vets, P.: 1989,
{\it Comm. Math. Phys.} {\bf 122},
pp.~249

\bibitem{Kuperberg2002}
Kuperberg G.: 2002,
{\it math-ph/0202035}

\bibitem{Michoel2003}
Michoel, T. and Nachtergaele, B.: 2003,
{\it math-ph/0310027}.

\bibitem{Sakurai}
Sakurai, J.J.: 1994,
'Modern Quantum Mechanics',
Addison-Wesley, Reading, Massachusetts

\bibitem{Linnik1971}
Ibargimov, I.A. and Linnik, Y.V.: 1971,
'Independent and Stationary Sequences of Random Variables',
Wolters-Noordhoff, Groningen/Netherlands

\bibitem{Billingsley1995}
Billingsley, P.: 1995,
'Probability and Measure',
John Wiley \& Sons, New York, 3rd ed.

\bibitem{Fick1990}
Fick, E. and Sauermann, G.: 1990,
'The Quantum Statistics of Dynamic Processes',
Springer , Berlin

\bibitem{Duerr1992}
D\"urr, D., Goldstein, S., Zanghi, N.: 2003,
{\it J. Stat. Phys.} {\bf 67} 
pp.~843

\bibitem{Duerr2003}
D\"urr, D., Goldstein, S., Zanghi, N.: 2004,
{\it J. Stat. Phys.} to be published,\\
{\it quant-ph/0308038}

\bibitem{Mahler1998}
Mahler, G. and Weberru\ss, V.: 2001,
'Quantum Networks',
Springer , Berlin

\bibitem{Jordan2003}
Jordan, A.N. and B\"uttiker, M.: 2003,
{\it cond-mat/0311647}

\bibitem{Hartmann2003}
Hartmann, M., Mahler, G. and Hess, O.:
{\it quant-ph/0312214}

\end{thebibliography}
\end{document}